\documentclass[seceq]{ptptex}
\usepackage{wrapft}
\usepackage{graphicx}

\title{%
Born-Again Braneworld
}

\author{
Sugumi \textsc{Kanno}$^{1,}
$\footnote{E-mail: kanno@phys.h.kyoto-u.ac.jp} ,
Misao \textsc{Sasaki}$^{2,}
$\footnote{E-mail: misao@vega.ess.sci.osaka-u.ac.jp}
and Jiro \textsc{Soda}$^{3,}
$\footnote{E-mail: jiro@phys.h.kyoto-u.ac.jp}
}

\inst{
$^1$Graduate School of Human and Environmental Studies, Kyoto University,\\
 Kyoto 606-8501, Japan\\
$^2$Department of Earth and Space Science, Graduate School of Science,\\ 
Osaka University, Toyonaka 560-0043, Japan\\
$^3$Department of Fundamental Sciences, FIHS, Kyoto University, \\
Kyoto 606-8501, Japan}

\recdate{
November 1, 2002}

\abst{%
We propose a cosmological braneworld scenario in which two branes
collide and emerge as reborn branes whose tensions have signs
opposite to the original tensions of the respective branes.
In this scenario, gravity on each of the branes is described by
a scalar-tensor-type theory in which the radion plays the role 
of the gravitational scalar, and the branes are assumed to be
inflating. However,  the whole dynamics is different 
from those of the usual inflation, due to the non-trivial dynamics
of the radion field. Transforming the conformal frame to
the Einstein frame, this born-again scenario resembles 
the pre-big-bang scenario. Thus, our scenario has features
of both inflation and pre-big-bang scenarios.
In particular, gravitational
waves produced from vacuum fluctuations may have a very
blue spectrum, while the inflaton field gives rise to
a standard scale-invariant spectrum.      
}

\begin{document}

\maketitle

\section{Introduction}

The inflationary universe scenario is a natural solution to
fundamental problems of the big-bang model, such as the horizon 
problem\cite{inflation}.
However, it is not a unique choice.  For example, a universe with an era
of contraction is also a possibility.
The pre-big-bang scenario is a realization of such a case in the
superstring context\cite{prebigbang}.
Unfortunately, however, the pre-big-bang scenario suffers from
the singularity problem, which cannot be solved without 
understanding the stringy non-perturbative effects.

One of the remarkable features of superstring theory is the existence
  of extra dimensions. Conventionally, the extra dimensions are considered
  to be compactified to form a small compact space of the Planck scale. 
However, recent revolutionary progress in string theory has lead to
the brane-world picture\cite{braneworld}.
 In this paper, we consider a system of two branes having tensions
of opposite sign, with the intermediate
spacetime (bulk) described by an anti-de Sitter space (AdS$_5$)\cite{RS1,radion1,radion2,cosmo}. 
 One of the branes is assumed to be our universe, and
there exists an inflaton field that leads to inflation.
The other brane is assumed to be vacuum,
but with a non-zero cosmological constant (see Fig.~1).

Assuming the slow roll of the inflaton field,
we can regard both branes as vacuum (de Sitter) branes.
Hence, we analyze this case in detail.
To this time, mostly the static de-Sitter two-brane system has been considered 
in the cosmological context\cite{dsbranes}.
 However, it is now well-known that
 a static de-Sitter two-brane system is unstable\cite{gen}.  
 We therefore investigate the non-trivial radion dynamics and
 focus on its cosmological consequences (See the previous work
 on the radion dynamics given in Ref.~10). As a result, we find 
 a new scenario of the braneworld, which we call
the ``born-again braneworld scenario". We show that the two branes 
can collide without developing serious singularities, as seen
from an observer on either brane, and emerge as reborn branes
with the signs of the Lagrangians reversed.
We find that our scenario has features common to
both the conventional inflationary scenario and the pre-big-bang
scenario. In a sense, we can regard it as a non-singular realization
of the pre-big-bang model in the braneworld context. (See related works
 and criticism of them presented in Refs.~11) and 12).) 
In particular, a flat spectrum for the density perturbation is
naturally produced, while background gravitational waves
with a very blue spectrum are generated through the collision.
It may be possible to detect this using future interferometric 
gravitational wave detectors\cite{detector}.

\begin{figure}
\centerline{\includegraphics[ width=6cm,height=5cm]{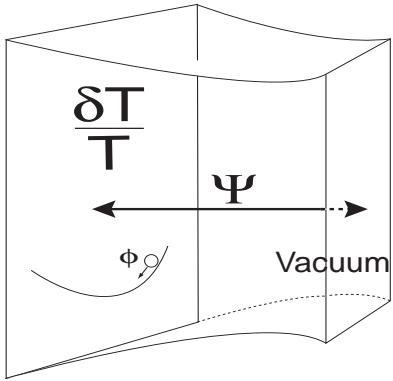}}
\caption{Radion as the distance between two branes.}
\end{figure}

\section{Effective action }

We begin by reviewing the effective equations on a brane
at low energies, which we derived in previous papers\cite{kanno1,kanno2}
 (see also Refs.~16) and 17)).
This effective theory is valid if the energy density on a brane
 is much smaller than the brane tension. Strictly speaking, we cannot
 use this action when two branes collide, because the junction conditions 
 lose their meaning in this case. Indeed, the collision process is singular 
 from
 the 5-dimensional point of view. However, this singularity is relatively 
 mild, and the action is completely regular at the  collision point. 
 This leads us to assume that the collision process can be described by 
 this effective action. This assumption is crucial for later analysis.     

Our system is described by the action 
\begin{eqnarray}
S&=&{1\over 2\kappa^2}\int d^5 x \sqrt{-g}\left({\cal R}
	+{12\over \ell^2}\right)
	-\sum_{i=A,B}\sigma_i \int d^4 x \sqrt{-g^{i\mathrm{\hbox{-}brane}}}
	\nonumber\\
&&	+\sum_{i=A,B} \int d^4 x \sqrt{-g^{i\mathrm{\hbox{-}brane}}}
\,{\cal L}_{\rm matter}^i \ ,
	\label{5D:action}
\end{eqnarray}
where ${\cal R}$, $g^{i\mathrm{\hbox{-}brane}}_{\mu\nu}$ and $\kappa^2$ are 
the 5-dimensional scalar curvature, the induced metric on the $i$-brane, 
and the 5-dimensional gravitational constant, respectively. We  consider 
an $S_1/Z_2$ orbifold spacetime with the two branes as the fixed points.
In the first Randall-Sundrum (RS1) model\cite{RS1},
the two flat 3-branes are embedded in AdS$_5$ with the curvature
radius $\ell$ and the brane tensions given by
$\sigma_A=6/(\kappa^2\ell)$ and $\sigma_B=-6/(\kappa^2\ell)$.
Then we have 
$g^{A\hbox{-}\rm brane}_{\mu\nu}=e^{2d/\ell}g^{B\hbox{-}\rm brane}$, 
where $d$ is the distance between the two branes.
We assume this model to be the ground state of our model.

Adding the energy momentum tensor to each of the two branes,
and allowing deviations from the pure AdS$_5$ bulk, the
effective (non-local) Einstein equations on the branes at low
energies take the forms\cite{ShiMaSa,kanno1,kanno2}
\begin{eqnarray}
G^{\mu}{}_{\nu} (h )
&=&{\kappa^2 \over\ell} T^{A\mu}{}_{\nu} 
	-{2\over\ell}\chi^{\mu}{}_{\nu} \,,
\label{A:einstein}
\\
G^{\mu}{}_{\nu}(f)
&=& -{\kappa^2 \over\ell} T^{B\mu}{}_{\nu} 
	-{2\over\ell} {\chi^{\mu}{}_{\nu} \over \Omega^4} \ .
\label{B:einstein}
\end{eqnarray}
where $h_{\mu\nu}=g^{A\hbox{-}{\rm brane}}_{\mu\nu}$
and $\Omega$ is a conformal factor that relates the metric
on the $A$-brane to that on the $B$-brane (specifically,
$f_{\mu\nu}=g^{B\hbox{-}{\rm brane}}_{\mu\nu}=\Omega^2h_{\mu\nu}$),
and the terms proportional to $\chi_{\mu\nu}$ are 5-dimensional
Weyl tensor contributions, which describe
the non-local 5-dimensional effect.
Although Eqs.~(\ref{A:einstein}) and (\ref{B:einstein})
are non-local individually, with undetermined $\chi_{\mu\nu}$,
they can be combined so as to reduce them to local equations
for each brane. Since $\chi_{\mu\nu}$ appears only algebraically,
one can easily eliminate $\chi_{\mu\nu}$ from Eqs.~(\ref{A:einstein}) 
and (\ref{B:einstein}). 

\subsection{$A$-brane}

First, consider the effective equations on the $A$-brane.
Defining a new field $\Psi = 1-\Omega^2$,  we find 
\begin{eqnarray}
G^{\mu}{}_{\nu}(h)&=&{\kappa^2 \over\ell \Psi } T^{A\mu}{}_{\nu}
	+{\kappa^2 (1-\Psi )^2 \over\ell\Psi} T^{B\mu}{}_{\nu}
	+{ 1 \over \Psi } \left(  \Psi^{|\mu}{}_{|\nu} 
  	-\delta^\mu_\nu  \Psi^{|\alpha}{}_{|\alpha} \right) \nonumber\\
&&	+{3 \over 2 \Psi (1-\Psi )} \left( \Psi^{|\mu}  \Psi_{|\nu}
  	- {1\over 2} \delta^\mu_\nu  \Psi^{|\alpha} \Psi_{|\alpha} 
  	\right),
  	\label{A:STG1} \\
\Box\Psi&=&{\kappa^2 \over 3\ell}(1-\Psi )
	\left\{ T^A + (1-\Psi)T^B  \right\}
	-{1 \over 2 (1-\Psi )} \Psi^{|\mu}\Psi_{|\mu} \ , 
  	\label{A:STG2}  	
\end{eqnarray}
where ``$|$" denotes the covariant derivative with respect to the metric
$h_{\mu\nu}$. Since $\Omega$ (or equivalently $\Psi$) contains
the information of the distance between the two branes,
we call $\Omega$ (or $\Psi$) the ``radion".

We can also determine $\chi^{\mu}{}_{\nu}$ by eliminating $G^{\mu}{}_{\nu}$ 
from Eqs.~(\ref{A:einstein}) and (\ref{B:einstein}). Then, we have
\begin{eqnarray}
\chi^{\mu}{}_{\nu}&=&-{\kappa^2(1-\Psi)\over 2 \Psi} 
	\left\{ T^{A\mu}{}_{\nu} 
	+ (1-\Psi)T^{B\mu}{}_{\nu}\right\}
	-{\ell\over 2 \Psi} \biggl[ \left(  \Psi^{|\mu}{}_{|\nu} 
	-\delta^\mu_\nu  \Psi^{|\alpha}{}_{|\alpha} \right) 
	\biggr. \nonumber \\
&&	\biggl.+{3 \over 2(1 -\Psi )} \left( \Psi^{|\mu}  \Psi_{|\nu}
  	-{1\over 2} \delta^\mu_\nu  \Psi^{|\alpha} \Psi_{|\alpha} 
  	\right) \biggr]   \ .  
  	\label{A:chi}
\end{eqnarray}
 Note that the index of $T^{B\mu}{}_{\nu}$ is to be raised
or lowered by the induced metric on the $B$-brane, $f_{\mu\nu}$.

The effective action for the $A$-brane that gives 
Eqs.~(\ref{A:STG1}) and (\ref{A:STG2}) is
\begin{eqnarray}
S_{\rm A}&=&{\ell\over 2 \kappa^2} \int d^4 x \sqrt{-h} 
	\left[ \Psi R (h) - {3 \over 2(1- \Psi )} 
     	\Psi^{|\alpha} \Psi_{|\alpha} \right] \nonumber\\
    &&\!\!\!\! 	+ \int d^4 x \sqrt{-h} {\cal L}^A 
      	+ \int d^4 x \sqrt{-h} \left(1-\Psi \right)^2 {\cal L}^B  
      	\ .  
      	\label{A:action} 
\end{eqnarray}

\subsection{$B$-brane} 

Using the same procedure as that above (but exchanging the roles of 
$h_{\mu\nu}$ and $f_{\mu\nu}$) also yields the effective equations on 
the $B$-brane.
Defining $\Phi = \Omega^{-2} -1$, 
we obtain
\begin{eqnarray}
G^\mu_{\ \nu}(f)&=&{\kappa^2 \over\ell\Phi }T^{B\mu}{}_{\nu}
	+{\kappa^2 (1+\Phi )^2 \over\ell\Phi} T^{A\mu}{}_{\nu}
	+{ 1 \over \Phi } \left(  \Phi^{;\mu}{}_{;\nu} 
  	-\delta^\mu_\nu  \Phi^{;\alpha}{}_{;\alpha} \right) \nonumber\\
&&	-{3 \over 2\Phi(1+\Phi)} \left( \Phi^{;\mu}  \Phi_{;\nu}
  	- {1\over 2} \delta^\mu_\nu  \Phi^{;\alpha} \Phi_{;\alpha} 
  	\right)  \ ,
  	\label{B:STG1} \\
\Box\Phi&=&{\kappa^2 \over 3\ell} (1+\Phi )
	\left\{ T^B + (1+\Phi) T^A \right\}
	+{1 \over 2 (1+\Phi )} \Phi^{;\mu} 
  	\Phi_{;\mu} \ .
  	\label{B:STG2} 
\end{eqnarray}
Here, ``$;$" denotes the covariant derivative with respect to the metric
$f_{\mu\nu}$. Note that the index of $T^{A\mu}{}_{\nu}$ is raised
or lowered by $h_{\mu\nu}$. Because $\Phi$ is equivalent to 
$\Omega$ or $\Psi$, we also call $\Phi$ the ``radion".

We can also express $\chi^{\mu}{}_{\nu}$ in terms of
quantities on the $B$-brane.
We find
\begin{eqnarray}
\chi^{\mu}{}_{\nu} &=& -{\kappa^2  \over 2 \Phi (1+\Phi)} 
	\left\{ T^{B\mu}{}_{\nu} 
	+ (1+ \Phi) T^{A\mu}{}_{\nu} \right\}  
	-{\ell\over 2 \Phi (1+\Phi)^2 } \biggl[ \biggl(  \Phi^{;\mu}{}_{;\nu} 
  	-\delta^\mu_\nu  \Phi^{;\alpha}{}_{;\alpha} \biggr) 
  	\biggr. \nonumber \\
&&	\biggl.-{3 \over 2(1+\Phi )} \left( \Phi^{;\mu}  \Phi_{;\nu}
  	- {1\over 2} \delta^\mu_\nu  \Phi^{;\alpha} \Phi_{;\alpha} 
  	\right) \biggr]   \ . 
  	\label{B:chi}
\end{eqnarray}

The effective action for the $B$-brane is given by 
\begin{eqnarray}
S_{\rm B}&=&{\ell\over 2 \kappa^2} \int d^4 x \sqrt{-f} 
	\left[ \Phi R(f) + {3 \over 2(1+\Phi )} 
     	\Phi^{;\alpha} \Phi_{;\alpha} \right]  \nonumber\\
&&\!\!\!\!	+\int d^4 x \sqrt{-f} {\cal L}^B
     	+\int d^4 x \sqrt{-f} {\cal L}^A(1+\Phi)^2
     	\ .  
      	\label{B:action} 
\end{eqnarray}

\section{Radion dynamics}

Most inflationary models are based on a slow-roll inflation
that has a sufficiently flat potential.
In this section, we consider the dynamics of branes with vacuum energy as  
a first-order approximation of a slow-roll inflation model.
Qualitative features of the brane cosmology can be understood with
this simplified vacuum brane model.

We take the matter Lagrangians to be ${\cal L}^A=-\delta\sigma^A$ and  
${\cal L}^B=-\delta\sigma^B$ in our effective action (\ref{A:action})
or (\ref{B:action}). 
The effective action on the $A$-brane in this case reads
\begin{eqnarray}
S_{\rm A} &=& {\ell\over 2 \kappa^2} \int d^4 x \sqrt{-h}
	\left[
	\Psi R
	- {3\over2(1-\Psi)}
	\Psi^{|\alpha}
	\Psi_{|\alpha} \right]
	-\delta\sigma^A \int d^4 x \sqrt{-h} \nonumber\\
&&	-\delta\sigma^B \int d^4 x \sqrt{-h} 
	\left(1-\Psi\right)^2. 
	\label{A:action-vcm}
\end{eqnarray}
Because our theory is a scalar-tensor-type theory, we call
this original action the ``Jordan-frame effective action".
 In order to study the dynamics of the radion, it is convenient to 
move to the Einstein frame, in which the action takes the canonical
Einstein-scalar form\cite{Nojiri}. 
Applying the conformal transformation defined by
$h_{\mu\nu}=\frac{1}{\Psi}g_{\mu\nu}$ and 
introducing the new field 
\begin{eqnarray}
\eta=-\log\left|\frac{\sqrt{1-\Psi}-1}{\sqrt{1-\Psi}+1}\right| \ ,
\label{A:field}
\end{eqnarray}
we obtain the Einstein-frame effective action as
\begin{eqnarray}
S_{\rm A}&=&{\ell\over 2 \kappa^2} \int d^4 x \sqrt{-g}
	\left[R(g)-\frac{3}{2}
	\nabla^\alpha\eta
	\nabla_\alpha\eta\right]
	-\int d^4x\sqrt{-g}~V(\eta)  \ ,
	\label{EH-action}
\end{eqnarray}
where $\nabla$ denotes the covariant derivative with respect to
 the metric $g_{\mu\nu}$, and the radion potential now takes the form
\begin{equation}
V(\eta) = \delta\sigma^A\left[~\cosh^4\frac{\eta}{2}
	+\beta\sinh^4\frac{\eta}{2}~\right], \quad 
	\beta=\frac{\delta\sigma^B}{\delta\sigma^A} \ .
	\label{potential}
\end{equation}

We can also start from the effective action 
on the $B$-brane to obtain the same Einstein-frame effective action.
By applying the conformal transformation defined by 
$f_{\mu\nu}=\frac{1}{\Phi}g_{\mu\nu}$ 
and introducing the new field
\begin{equation}
\eta=-\log\left|\frac{\sqrt{\Phi+1}-1}{\sqrt{\Phi+1}+1}\right|  \ ,
\label{B:field}
\end{equation}
we also arrive at Eq.~(\ref{EH-action}). 

We are now ready to examine the radion dynamics (see Fig.~2).
\begin{figure}[h]
\center{\includegraphics[height=5cm, width=6cm]{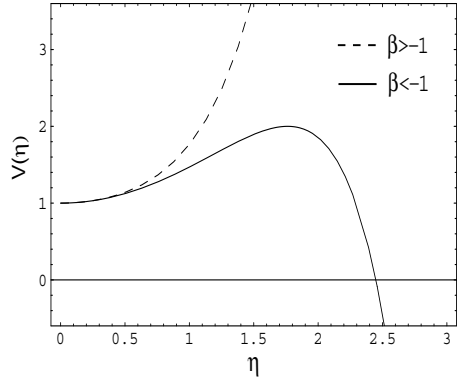}}
\caption{The potential function for $\eta$. }
\end{figure}
Notice that the two branes are infinitely separated when $\eta=0~(\Psi=1)$,
and they collide when $\eta=\infty~(\Psi=0)$. 
For definiteness, let us assume $\delta\sigma_A>0$. 
If $\delta\sigma^A+\delta\sigma^B>0$, $\Psi$ will
move towards unity; i.e., the branes will move away from each other.
If $\delta\sigma^A+\delta\sigma^B<0$,
the potential has a maximum at $\Psi_c=1+1/\beta$, and the behavior
depends on whether $\Psi>\Psi_c$ or $\Psi<\Psi_c$. 
If $\Psi>\Psi_c$, the branes will become infinitely separated.
If $\Psi<\Psi_c$, the branes will approach each other
and eventually collide. 

The static two de-Sitter brane solution corresponds to 
the unstable point ~$\Psi=\Psi_c$.
In fact, considering the fluctuations around $\Psi_c$,
 we find an instability characterized by the equation
\begin{equation}
\delta \ddot{\Psi} + 3 H \delta \dot{\Psi}
 -4\left(H^2+\frac{K}{a^2}\right)\delta \Psi=0   \ .
\label{perturbation}
\end{equation}
We see that the mass square, $-4(H^2+K/a^2)$, is negative,
in accordance with the previous linear perturbation
 analysis\cite{gen}.

As we mentioned above, in the case $\Psi<\Psi_c$,
the two branes collide. From the 5-dimensional point of view,
this is certainly a singularity, where the spacetime degenerates
to 4 dimensions. However, as far as observers on the branes
are concerned, nothing seems to go wrong.
In fact, the action~(\ref{A:action}) is
well-defined even in the limit $\Psi\to0$.
Let us assume that $\Psi$ smoothly becomes negative after collision.
Then replacing $\Psi$ as $\Psi\to -\tilde{\Psi}$ in the
action (\ref{A:action}), we find
\begin{eqnarray}
-S_A &=& {\ell\over 2\kappa^2} \int d^4 x \sqrt{-h}
	\left[~
	\tilde{\Psi}R(h)
	+\frac{3}{2}\frac{1}{1+\tilde{\Psi}}
	\tilde{\Psi}^{|\alpha}\tilde{\Psi}_{|\alpha}~
	\right]
	+\int d^4 x \sqrt{-h}
	(-{\cal L}^A) \nonumber \\
&&	+\int d^4 x \sqrt{-h} 
	\left(1+\tilde{\Psi}\right)^2
	({-\cal L}^B) \ .
\end{eqnarray}
This is the same as the effective action on the $B$-brane, given in
Eq.~(\ref{B:action}), except for the overall change of sign and
the associated changes of sign of the matter Lagrangians.
This fact can be interpreted as follows. After collision,
the positive tension brane becomes a negative tension brane,
together with the sign change of the matter Lagrangian, and
vice versa for the initially negative tension brane.
This implies that, if we live on either of the branes,
our world transmutes into quite a different world, and 
so do we without much damage to the world.
 That is, we are born again!
 
This procedure might cause a serious problem when we consider 
 quantum theory. A similar issue arises in string theory
 if there exists a negative tension brane. However, string theory
 has the potential ability to overcome this difficulty.  For the time being, 
 we can only hope that our prescription has an appropriate interpretation
 in the context of string theory.  
 
\section{Born-again braneworld}

After the collision, if our world had initially been a positive tension brane,
we would now be on the negative tension brane.
However, the theory described by the action (\ref{B:action})
with any value of $\Phi$ contradicts observation.
Therefore we assume that we were initially on the
negative tension brane ($B$-brane) before the collision. 

Let us first investigate the cosmological evolution of the
$B$-brane in the original Jordan frame.
We consider the spatially isotropic and homogeneous metric on
the brane
\begin{eqnarray}
ds^2=-dt^2+a^2(t)\gamma_{ij}dx^idx^j\,,
\label{mtrc:cosmology}
\end{eqnarray}
where $a(t)$ is the scale factor and $\gamma_{ij}$ is the metric of a
maximally symmetric 3-space with comoving curvature $K=0$, $\pm1$.
Using Eqs.~(\ref{B:STG1}) and (\ref{B:STG2}), the field
equations on the $B$-brane can be written
\begin{eqnarray}
-3\left(H^2+\frac{K}{a^2}\right)
&=&-\frac{\kappa^2}{\ell}\frac{1}{\Phi}\delta\sigma^B
	-\frac{\kappa^2}{\ell}\frac{(1+\Phi)^2}{\Phi}\delta\sigma^A
	+3H\frac{\dot{\Phi}}{\Phi}
	+\frac{3}{4}\frac{\dot{\Phi}^2}{\Phi(1+\Phi)} \ ,
	\label{B:00} \\
-2\left(\dot{H}-\frac{K}{a^2}\right)
&-&3\left(H^2+\frac{K}{a^2}\right)
\nonumber\\
&&\hspace{-1cm}
	=-\frac{\kappa^2}{\ell}\frac{1}{\Phi}\delta\sigma^B 
	-\frac{\kappa^2}{\ell}\frac{(1+\Phi)^2}{\Phi}\delta\sigma^A
	+\frac{\ddot{\Phi}}{\Phi}
	+2H \frac{\dot{\Phi}}{\Phi}
	-\frac{3}{4}\frac{\dot{\Phi}^2}{\Phi(1+\Phi)} 
	\label{B:ij} \ , \\
\ddot{\Phi}+3H\dot{\Phi}
&=&\frac{4\kappa^2}{3\ell}(1+\Phi)
	\left[\delta\sigma^B+(1+\Phi)\delta\sigma^A\right]
	+\frac{1}{2}\frac{1}{1+\Phi}\dot{\Phi}^2 \ .
	\label{B:radion}
\end{eqnarray}
Note that Eq.~(\ref{B:00}) is the Hamiltonian constraint. 
Eliminating $\ddot{\Phi}$ from Eq.~(\ref{B:ij}) by using Eq.~(\ref{B:radion}) 
and combining the resulting equation with Eq.~(\ref{B:00}),  
we obtain 
\begin{eqnarray}
\dot{H}-\frac{K}{a^2}=-2\left(H^2+\frac{K}{a^2}\right)
-\frac{2\kappa^2}{3\ell}\delta\sigma^B \ .
\label{B:ij2} 
\end{eqnarray}
Integrating this equation, we obtain the Friedmann equation 
with dark radiation,
\begin{eqnarray}
H^2+\frac{K}{a^2}
=-\frac{\kappa^2}{3\ell}\delta\sigma^B+\frac{C}{a^4} \ .
\label{friedmann}
\end{eqnarray}
Note that this is just a cosmological version of the non-local
Einstein equations on the brane, given in Eq.~(\ref{B:einstein}), in which
the $\chi^{\mu}{}_\nu$ term gives the dark radiation $C/a^4$.

Comparing Eq.~(\ref{B:00}) with Eq.~(\ref{friedmann}), 
we find the following relation between the radion 
and the dark radiation:
\begin{eqnarray}
&&\frac{\kappa^2\delta\sigma^B}{3\ell}\frac{1+\Phi}{\Phi}
	\left[1+\frac{(1+\Phi)}{\beta}\right]  
	-H\frac{\dot{\Phi}}{\Phi}
	-\frac{1}{4}\frac{1}{1+\Phi}\frac{\dot{\Phi}^2}{\Phi}
	=\frac{C}{a^4}   \ . 
\label{relation}
\end{eqnarray}
This gives, in particular, the relation between the
initial conditions of the radion and the sign of the dark 
radiation.

Assuming $\delta\sigma_B<0$, the Friedmann equation (\ref{friedmann})
yields the dependence of $H$ on the dark radiation.
Setting $H_*^2=(\kappa^2/3\ell)(-\delta\sigma_B)$,
we find
\begin{eqnarray}
H_*^2K^2&-&4C>0\,:
\nonumber\\
H&=&H_*\frac{\sqrt{H_*^2K^2-4C}\sinh2H_* t}
{\sqrt{H_*^2K^2-4C}\cosh2H_* t+H_*^2K} \ ,
\\
~\nonumber\\
H_*^2K^2&-&4C=0\,:
\nonumber\\
H&=&H_*\frac{2\,e^{2H_*t}}{2\,e^{2H_*t}+H_*^2K} \ ,
\\
~\nonumber\\
H_*^2K^2&-&4C<0\,:
\nonumber\\
H&=&H_*\frac{\sqrt{4C-H_*^2K^2}\cosh2H_* t}
	{\sqrt{4C-H_*^2K^2}\sinh2H_* t+H_*^2K} \ .
\end{eqnarray}

To realize the born-again braneworld scenario,
we consider the case of colliding branes. For simplicity,
we assume $K=0$. A numerical solution of $\Phi$ is
displayed in Fig.~3. We indeed see that $\Phi$ passes through
zero smoothly and approaches $-1$; i.e.,
the reborn branes will eventually be infinitely separated.
\begin{figure}[h]
\center{\includegraphics[height=5cm, width=6cm]{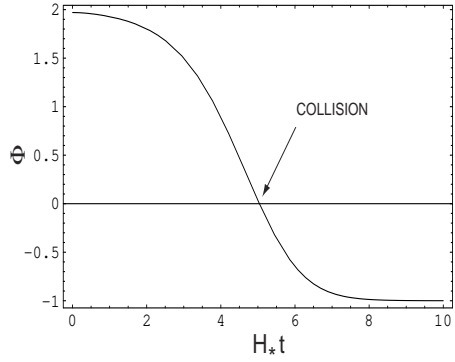}}
\caption{The time evolution of $\Phi$. The evolution is completely
 regular at the collision point. }
\end{figure}
\begin{figure}[h]
\center{\includegraphics[height=5cm, width=6cm]{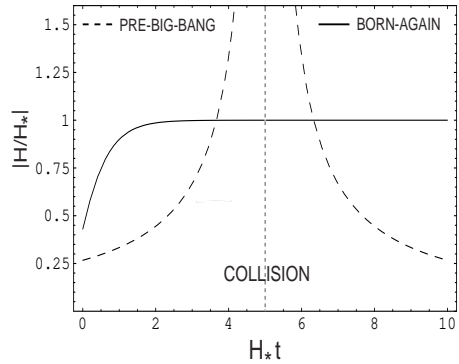}}
\caption{The evolution of the Hubble constant in the Jordan frame.
 The solution rapidly approaches the de-Sitter spacetime. 
 We also plotted the pre-big-bang solution in the Einstein frame.}
\end{figure}

Let us analyze this collision. 
We denote the Hubble constant at the time of collision $t=t_c$ by $H_c$. 
Applying Eq.~(\ref{relation}) to the vicinity of the time of collision,
we find
\begin{eqnarray}
\Phi=-2(1-\sqrt{\gamma})H_c(t-t_c)\,;
\quad\gamma=1-\frac{H_*^2}{H_c^2}\left(1+\frac{1}{\beta}\right)\,.
\end{eqnarray}
As expected, $\Phi$ behaves perfectly smoothly around
the time of collision. The brane geometry is, of course,
perfectly regular as well. In fact, the Friedmann
equation (\ref{friedmann}) continues to hold without
a hint of collision.

Now, we transform these quantities into the Einstein frame.
Because $\Phi\to-1$ eventually, we can regard our present
universe to be described by the Einstein frame.
The relation between the Einstein frame and the Jordan frame is
\begin{eqnarray}
ds^2_E&=&-dt_E^2 + b^2 (t_E) \delta_{ij} dx^i dx^j \nonumber\\
	&=&|\Phi| \left[-dt_J^2+a(t_J)^2\delta_{ij}dx^idx^j\right],
\end{eqnarray}
where we have attached the subscripts $E$ and $J$ to the time coordinates
to denote the cosmic time in the Einstein frame and the Jordan frame,
respectively.
Thus we have
\begin{eqnarray}
b=\sqrt{|\Phi|}\,a, \quad dt_E = \sqrt{|\Phi|}\,dt_J \ .
\label{conftrans}
\end{eqnarray}
Therefore, the Hubble parameter in the Einstein frame 
behaves in the vicinity of collision as
\begin{eqnarray}
\frac{\dot{b}(t_E)}{b(t_E)}
=\frac{1}{3t_E} 
+\frac{H_c}{\left(3(1-\sqrt{\gamma})H_c|t_E|\right)^{1/3}}\,,
\label{Ehubble}
\end{eqnarray}
where the collision time in the Einstein frame is set to be $t_E=0$.

We note that in the Einstein frame, the universe contracts
rapidly just before the collision, and the Hubble parameter 
diverges to $-\infty$ at collision.
Then, the universe is reborn with an infinitely large Hubble parameter,
which looks like a big-bang singularity. Thus, because there exists
no singularity in the Jordan frame, the pre-big-bang phase and the
post-big-bang phase in the Einstein frame are successfully
connected. That is, our scenario is indeed a successful realization
of the pre-big-bang scenario in the context of the braneworld (see Fig.~4).

\section{Observational implication}

 As we can see from Eq.~(\ref{friedmann}), the universe will rapidly
 converge to the quasi-de-Sitter regime, while the radion can 
 vary, as long as the relation (\ref{relation}) is satisfied. 
 In the Jordan frame, because the metric couples with the radion,
 the non-trivial evolution of the radion field affects the perturbations.
 This possibility discriminates our model from the usual inflationary 
 scenario. On the other hand, the inflaton does not couple directly
with the radion field. Hence, the inflaton fluctuations are expected
to give adiabatic fluctuations with a flat spectrum. 
 This feature of our model is an advantage it has over the pre-big-bang model.

\subsection{Radion fluctuations}

 To study the behavior of the radion fluctuations,
 it is convenient to work in the Einstein frame. 
We express the metric perturbation in the Einstein frame as
\begin{eqnarray}
ds_E^2
&=&b^2\left[-(1+2A)d\tau^2+2\partial_iBdx^id\tau\right.
\left.+\left((1+2{\cal R})\delta_{ij}+2\partial_i\partial_jE\right)
dx^idx^j\right]
\label{mtrc:sclr-ptb}
\end{eqnarray}
The action for a curvature perturbation ${\cal R}$
on the $\delta\eta=0$ (i.e., radion-comoving) slice reads
(for a concise review, see Appendix B of Ref.~20))
\begin{eqnarray}
S={1\over2}\int d\eta\, d^3x\,z^2
\left[{\cal R}_c'{}^2-{\cal R}_c^{\,|i}{\cal R}_{c\,|i}\right]\,,
\label{ptb:action}
\end{eqnarray}
where ${\cal H}=b'/b$ and
\begin{eqnarray}
{\cal R}_c={\cal R}-{\cal H}{\delta\eta\over\eta'}\,,
\quad
z=\sqrt{3\ell\over2\kappa^2}\,{b\eta'\over{\cal H}}\,.
\label{Rcdef}
\end{eqnarray}
The equation of motion for ${\cal R}_c$ is
\begin{equation}
      {\cal R}_c''+ 2{z'\over z}{\cal R}_c'
-\mathop\Delta^{(3)}{\cal R}_c = 0 \ .
\label{sclr:schrodinger-typ}
\end{equation}

Because the background behaves as $b \sim (-\tau)^{1/2} $,
${\cal H}\sim(2\tau)^{-1}$ and $\eta' \sim (-\tau)^{-1}$, 
we have $z\propto b$, and
the positive frequency modes for the adiabatic vacuum are given by
\begin{equation}
   {\cal R}_{c,k}\sim\sqrt{\pi\kappa^2\over6H_*\ell}\,
H_0^{(1)} (-k\tau ) \,,
\label{bessel}
\end{equation}
where we have normalized $b$ as $b=|H_*\tau|^{1/2}$.
Then we have
\begin{eqnarray}
\left\langle{\cal R}_c^2\right\rangle_k
\equiv{k^3\over 2\pi^2}P(k)
={k^3\over 2\pi^2}|{\cal R}_{c,k}|^2
\sim {k^3\over H_*M_{pl}^2}\,,
\label{curvamp}
\end{eqnarray}
where $M_{pl}^2=\kappa^2/\ell$. Thus the spectrum is very blue.
If we define the spectral index by $P(k)\propto k^{n-4}$,
this implies $n=4$.

There are, however, a couple of points that should be mentioned.
First, the curvature perturbation ${\cal R}_c$ is logarithmically
divergent at the instant of collision. This divergence does not
disappear even in the Jordan frame.\footnote{
The fact that the perturbation is generically singular
when the effective gravitational constant changes sign
was first pointed out by Starobinsky\cite{star}.}. Note that
${\cal R}_c$ is defined on the hypersurface on which the radion
is uniform, and hence is invariant under the conformal transformation
(\ref{conftrans})
This suggests the marginal instability of our system.
Nevertheless, if we cut off an infinitesimally small time interval
around the singularity, say $[-\epsilon,\epsilon]$,
and smoothly match ${\cal R}_c$ at $\tau=\pm\epsilon$,
the result is quite insensitive to the choice of $\epsilon$,
as long as $k\epsilon\ll1$, i.e., the scale is outside the
effective Hubble horizon ${\cal H}^{-1}$.
Thus, we expect the result (\ref{curvamp}) to be valid
for all scales of cosmological interest.

The second point, which is a possible drawback,
is the following. If inflation (in the Jordan frame) 
continues for a time of $O(H_*^{-1})$ after collision,
the blue spectrum given above should not be called ``blue" after all.
As can be seen from Fig.~4, or from Eq.~(\ref{Ehubble}),
$|\cal H|$ is quite symmetric around $\tau=0$, at least 
for $|\tau|\lesssim H_*^{-1}$. This implies that all the modes
with $k>H_*$ that were once outside the horizon in the 
pre-big-bang phase came inside the horizon again in the
post-big-bang phase by the time $\tau\sim H_*^{-1}$.
(It should be noted that the conformal time is approximately
equal to the cosmic time in the Jordan frame near the collision,
with our normalization such that $b=|H_*\tau|^{1/2}$.)
Because the evolution near the collision time is approximately
time symmetric, the standard vacuum state will re-emerge
for the modes that re-enter the horizon in the Einstein frame.
These modes will then come outside the horizon again
in the inflationary phase, and their spectrum
will be a normal scale-invariant one.
Therefore, we would obtain a spectrum that is
scale-invariant for $k>H_*$ and blue spectrum
at $k<H_*$, with the maximum amplitude given by
\begin{eqnarray}
\left\langle{\cal R}_c^2\right\rangle_{k=H_*}
\sim {H_*^2\over M_{pl}^2}\,.
\end{eqnarray}
For the values of $H_*$ predicted by the standard inflationary models,
this is not really a blue spectrum in the observational sense.

This problem can be avoided only if the inflation ends
right after the collision, when $\tau\ll H_*^{-1}$.
One possibility is to resort to the marginal
instability mentioned above. There may be a model in which
the marginally divergent spectrum at high frequencies
triggers a phase transition to end inflation.
It is not clear if it is possible to construct such a model
in a natural way. We leave investigation of this point as a future problem.

\subsection{Gravitational waves}

Next, consider the tensor perturbations
\begin{equation}
ds^2 = b^2 (\tau) \left[
       -d\tau^2 + (\delta_{ij} + h_{ij}) dx^i dx^j \right] \ , 
\label{mtrc:tnsr-ptb}
\end{equation}
where $h_{ij}$ satisfy the transverse-traceless conditions 
 $h_{ij}{}^{,j} = h^{i}{}_i =0$. 
For the gravitational tensor perturbations, we have
\begin{equation}
    h_k^{\prime\prime} + 2{\cal H}h_k'+k^2h_k
    =0 \ ,
\label{tnsr:schrodinger-typ}
\end{equation}
where $h_k$ is the amplitude of $h_{ij}$.
Since ${\cal H}\sim (2\tau)^{-1}$,
$h_k$ has approximately the same spectrum as ${\cal R}_c$,
including the magnitude.
In particular, the spectral index for the gravitational waves is also
 $n=4$. (Here, the spectral index is defined by $P_h(k)\propto k^{n-4}$, as in
the case of the scalar curvature perturbation. For the tensor perturbation,
the conventional definition is ~$n_T=n-1$.)
 Provided that inflation ends right after collision,
as discussed in the previous subsection, this gives a sufficiently
blue spectrum that can amplify $\Omega_g$ by several orders of 
magnitude or more on small scales as compared to conventional
inflation models. Thus, there is the possibility that
it may be detected by a space laser interferometer for
low frequency gravitational waves, such as LISA~\cite{detector}.
  
\subsection{Inflaton perturbation}

To investigate the inflaton perturbation rigorously,
one needs to introduce an inflaton field explicitly and
consider a system of equations fully coupled with
the radion and the metric perturbation. However, since
this is beyond the scope of the present paper, let us
just estimate the effect of the metric perturbation
induced by radion fluctuations on the inflaton perturbation.

The field equation for the inflaton, $\varphi$, 
in the Jordan frame is
\begin{eqnarray}
&&a^{-2}(a^2\delta\varphi')'
-\Delta\delta\varphi+a^2\partial_\varphi^2V\delta\varphi
\nonumber\\
&&\quad\qquad=-3\varphi'{\cal R}_J'+a^{-2}(a^2\varphi'A_J)'
-a^2\partial_\varphi VA_J
-\Delta(E_J'-B_J)\varphi'\,,
\label{Feq}
\end{eqnarray}
where the suffix $J$ indicates a quantity in the Jordan frame.
Again, let us consider the radion-comoving slice. 
Then, all the metric perturbation variables in the Jordan frame
coincide with their respective counterparts in the Einstein frame.

Ignoring the effect of the inflaton perturbation,
the Hamiltonian and momentum constraints on the
radion-comoving slice are~\cite{KS}
\begin{eqnarray}
\Delta({\cal R}_c-{\cal H}(E'-B)_c)
&=&{3\over4}\eta'{}^2A_c \ ,
\nonumber\\
{\cal R}_c'-{\cal H}A_c
&=&0 \ .
\label{Econstraints}
\end{eqnarray}
Using these equations, Eq.~(\ref{Feq}) reduces to
\begin{eqnarray}
&&a^{-2}(a^2\delta\varphi')'
-\Delta\delta\varphi+a^2\partial_\varphi^2V\delta\varphi
\nonumber\\
&&\qquad
=-{2{\cal H}^2+{\cal H}'\over{\cal H}^2}{\cal R}_c'\varphi'
+a^{-2}\left(a^2{{\cal R}_c'\over{\cal H}}\varphi'\right)'
-{{\cal R}_c'\over{\cal H}}a^2\partial_\varphi V
-{\Delta{\cal R}_c\over{\cal H}}{\varphi'}\,.
\end{eqnarray}
 Since ${\cal R}_c'/{\cal H}$ is finite,
we see that the right-hand side of the above is regular and small
for the slow-roll inflation. Hence, the inflaton 
fluctuations are not strongly affected by the radion fluctuations.
Thus, the inflaton fluctuations should have a standard
scale-invariant spectrum.

\section{Conclusion}

 In this paper, we proposed a scenario in which two branes collide
and are reborn as new branes, called the ``born-again braneworld scenario".
 Our model has the features of both inflationary and pre-big-bang scenarios. 
In the original frame, which we call the Jordan frame, because
gravity on the brane is described by a scalar-tensor-type
theory, the brane universe is assumed to be inflating due to an
  inflaton potential. 
From the 5-dimensional point of view, the radion, which represents the 
distance between the branes
and which acts as a gravitational scalar on the branes,
 has non-trivial dynamics and these vacuum branes can collide and 
    pass through smoothly.
After collision, it is found that 
the positive tension and the negative tension 
 branes exchange their role.  Then, they move away from each other,
and the radion becomes trivial after a sufficient lapse of time.
The gravity on the originally negative tension
brane (whose tension becomes positive after collision) then
approaches that of the conventional Einstein theory, except for 
tiny Kaluza-Klein corrections. 

We can also consider the cosmological evolution of the branes in
the Einstein frame. Note that the two frames are indistinguishable
at present if our universe is on the positive tension brane after
collision. In the Einstein frame, the brane universe is contracting
before the collision and a singularity is encountered 
at the collision point. This resembles the pre-big-bang scenario.
Thus our scenario may be regarded as a non-singular realization
of the pre-big-bang scenario in the braneworld context.   
 
Because our braneworld is inflating, and the inflaton has essentially
no coupling with the radion field, 
an adiabatic density perturbation with a flat spectrum is 
naturally realized.
  On the other hand, because the collision of branes mimics the pre-big-bang 
  scenario,
the primordial background gravitational waves 
with a very blue spectrum may be produced.
This suggests the possibility that we may be able to observe 
the collision epoch using a future gravitational wave
detector, such as LISA. 

Admittedly, the collision process must be treated with
a more fundamental theory.
However, because the singularity at the collision point is
very mild, we expect that the qualitative features
of our scenario will remain unchanged, even if
we include the effect of a (yet unknown) fundamental theory.
 The born-again scenario surely deserves further investigation.

\section*{Acknowledgements}
This work was supported in part by Monbukagaku-sho Grant-in-Aid 
for Scientific Research, Nos.~14540258, 1047214 and 12640269.
We would like to thank D. Langlois and C. van de Bruck
 for discussions and comments.
A part of this work was done while one of us (MS) was visiting
the gravitation and cosmology group (GRECO) at IAP in Paris. 
MS would like to thank the GRECO members of IAP
for warm hospitality.

\end{document}